\documentclass[bbm,amsfonts,epsfig,12pt,here]{article}
\usepackage{graphicx}
\usepackage{dcolumn}
\usepackage{bm}

\newcommand{\drawsquare}[2]{\hbox{%
\rule{#2pt}{#1pt}\hskip-#2pt
\rule{#1pt}{#2pt}\hskip-#1pt
\rule[#1pt]{#1pt}{#2pt}}\rule[#1pt]{#2pt}{#2pt}\hskip-#2pt
\rule{#2pt}{#1pt}}

\newcommand{\Yfund}{\raisebox{-.5pt}{\drawsquare{6.5}{0.4}}}
\newcommand{\Ysymm}{\Yfund\hskip-0.4pt%
                    \Yfund}
\def\symm{\Ysymm}
\def\bsymm{\overline{\Ysymm}}

\def\drawbox#1#2{\hrule height#2pt
        \hbox{\vrule width#2pt height#1pt \kern#1pt
              \vrule width#2pt}
              \hrule height#2pt}

\def\Asym#1#2{\vcenter{\vbox{\drawbox{#1}{#2}
              \kern-#2pt       
              \drawbox{#1}{#2}}}}

\def\asymm{\Asym{6.4}{0.3}}

\def\basymm{\overline{\asymm}}

\def\Acknowledgements{\bigskip  \bigskip {\begin{center}
              \bf Acknowledgments \end{center}}}

\newcommand {\beq} {\begin{equation}}
\newcommand {\eeq} {\end{equation}}
 \newcommand{\be}{\begin{eqnarray}}
\newcommand{\ee}{\end{eqnarray}}

\begin{document}

\begin{titlepage}

\begin{flushright}
\small FTPI-MINN-03/26, UMN-TH-2215/03 \\
NORDITA-2003-60HE
\end{flushright}

\vspace {1cm}

\centerline{{\Large \bf Effective Lagrangians for Orientifold Theories}}

\vskip 1cm \centerline{\large F. Sannino ${}^a$ and M. Shifman
${}^{b}$} \vskip 0.1cm

\vskip 0.5cm
\centerline{${}^a$  NORDITA \& The Niels Bohr Institute, Blegdamsvej 17,}
\centerline{ DK-2100 Copenhagen \O, Denmark }
\vskip 0.5cm
\centerline{${}^b$ William I. Fine Theoretical Physics Institute,
University
of Minnesota,}
\centerline{Minneapolis, MN 55455, USA}

\vskip 1cm

\begin{abstract}
We construct  effective Lagrangians of the Veneziano-Yankielowicz
(VY) type for two non-supersymmetric theories which are
orientifold daughters of supersymmetric gluodynamics (containing
one Dirac fer\-mion in the two-index antisymmetric or symmetric
representation of the gauge group). Since the parent and daughter
theories are planar equivalent, at $N\to\infty$ the effective
Lagrangians in the orientifold theories basically coincide with
the bosonic part of the VY Lagrangian.

We depart from the supersymmetric limit in two ways. First, we
consider finite (albeit large) values of $N$. Then $1/N$ effects
break supersymmetry. We suggest seemingly the simplest
modification of the VY Lagrangian which incorporates these $1/N$
effects, leading to a non-vanishing vacuum
energy density. We analyze the spectrum of the finite-$N$
non-supersymmetric daughters. For $N=3$ the two-index
antisymmetric representation (one flavor)
is equivalent to one-flavor QCD. We
show that in this case the scalar quark-antiquark state is heavier than the
corresponding pseudoscalar state,  `` $\eta^{\prime}$''.
Second, we add a small fermion mass term. The fermion mass term  breaks
supersymmetry explicitly.  The
 vacuum degeneracy is lifted. The parity doublets split.
We evaluate the splitting.
Finally, we  include the $\theta$-angle and study its implications.

\end{abstract}
\end{titlepage}

\section{Introduction}
\label{uno}

Recently it has been argued \cite{Armoni:2003gp,Armoni:2003fb}
that {\em non}-supersymmetric Yang-Mills theories with the fermion
sector presented  in Table~\ref{table}  are nonperturbatively
equivalent to supersymmetric Yang-Mills (SYM) theory at large $N$,
so that exact results established in SYM theory (e.g.
\cite{{Shifman:1999mv},{Shifman:ia}}) should hold also in
these ``orientifold" theories. For example, the orientifold
theories, at large $N$, must have  an exactly calculable bifermion
condensate and an infinite number of degeneracies  in the spectrum
of color-singlet hadrons. (Needless to say, both theories, the
SUSY parent and orientifold daughter, are expected to confine.)
The phenomenon goes under the name of planar equivalence; it does
not mean, however, the full parent-daughter coincidence. For
instance, at $N\rightarrow \infty$ the color-singlet spectrum of
the orientifold theories does not include composite fermions. The
planar equivalence relates corresponding {\em bosonic} sectors in
the corresponding vacua of two theories. Some predictions for
one-flavor QCD (which is the antisymmetric orientifold daughter at
$N=3$) were made along these lines in \cite{Armoni:2003fb,ASV3}.
\begin{table}[h]
\begin{center}
\begin{minipage}{2.8in}
\begin{tabular}{c||ccc }
 & $SU(N)$ & $U_V(1)$ & $U_A(1)$  \\
  \hline \hline \\
${\psi_{\{ij\}}}$& $\symm$ & $1$ & $1$  \\
 &&&\\
 $\widetilde{\psi}^{\{ij\}}$ &$\bsymm $& $-1$ & $1$ \\
 &&&\\
$G_{\mu}$ &{\rm Adj} & $0$ & $0$    \\
\end{tabular}
\end{minipage}
\begin{minipage}{2.2 in}
\begin{tabular}{c||ccc}
 & $SU(N)$ & $U_V(1)$ & $U_A(1)$  \\
  \hline \hline \\
$\psi_{[ij]}$& $\asymm$ & $1$ & $1$  \\
 &&&\\
 $\widetilde{\psi}^{[ij]}$ &$\basymm $& $-1$ & $1$ \\
 &&&\\
$G_{\mu}$ &{\rm Adj} & $0$ & $0$    \\
\end{tabular}
\end{minipage}
\end{center}
\caption{The fermion sector of the orientifold  theories. $\psi$
and $\widetilde{\psi}$ are two Weyl fermions, while $G_{\mu}$
stands for the gauge bosons. In the left (right) parts of the
table the fermions are in the two-index symmetric (antisymmetric)
representation of the gauge group SU$(N)$. $U_V(1)$ is the
conserved global symmetry while the $U_A(1)$ symmetry is lost at
the quantum level due to the chiral anomaly.} \label{table}
\end{table}

\noindent
The name of {\it orientifold field theory} is borrowed from
string-theory terminology. In fact, these theories were shown to
live on a brane configuration of type 0A string theory
\cite{Armoni:1999gc,Angelantonj:1999qg} which consists of NS5
branes, D4 branes and an {\it orientifold} plane. The gauge groups
in the parent and daughter theories are the same, and so are the
gauge couplings.

In this paper we construct effective Lagrangians for orientifold
theories in terms of ``important" low-lying color-singlet states.
The effective Lagrangians of this type have a long history
\cite{schechter,{joe},{MS},{SST},{Veneziano:1982ah}} and are known
to concisely encode nonperturbative aspects of strongly coupled
theories, such as the vacuum structure and symmetries, both exact
and anomalous. Among recent developments in this direction was the
demonstration of how the information on the center of the SU($N$)
gauge group (i.e.  $Z_N$) is efficiently transferred to the
hadronic states \cite{Sannino:2002wb}. This demonstration led to a
deeper understanding of the deconfining phase transition
\cite{Mocsy:2003tr} in pure Yang-Mills theory. When quarks were
added, either in the fundamental or in the adjoint representations
of the gauge group, a link between the chiral and deconfining
phase transitions was uncovered~\cite{Mocsy:2003qw}.

Similarly, the construction of the effective Lagrangians for the
orientifold field theories will allow us to economically describe
the vacuum structure of the theory in the $N\rightarrow\infty$
limit, and, more importantly, will provide us with a tool to
investigate effects due to finite $N$ and due to a nonvanishing
fermion mass, such as the scalar-pseudoscalar
splitting. This is the main focus of the present work.

We mostly investigate the effective Lagrangian for the orientifold
theory with fermions in the two-index antisymmetric representation
of the gauge group. Distinctions for the fermions in the two-index
symmetric representation are briefly summarized in Sect.~\ref{tisym}.
The former theory
is of particular interest since at $N=3$ it is nothing but QCD
with one quark flavor \cite{Armoni:2003fb}. The effective
Lagrangians for the fermions in the two-index symmetric or
antisymmetric representations
are identical at $N=\infty$; differences emerge as
subleading $1/N$ effects at finite $N$. In Sect.~\ref{due} we
review the SYM effective Lagrangian. We construct the effective
Lagrangians for the orientifold theories at large $N$ in
Sect.~\ref{tre}. In the limit   $N\rightarrow \infty$
they reduce to the bosonic part of the effective Lagrangian in the
supersymmetric parent theory. The
vanishing of the vacuum energy, parity doubling of opposite-parity
states and the $N$-fold vacuum degeneracy characterize the large
$N$ limit \cite{Armoni:2003fb} of the orientifold theories we
investigate. We then construct the effective Lagrangian for the
two-index antisymmetric representation at finite $N$ (Sect.~\ref{tre-finite-N}).
The effective Lagrangian encodes the correct large $N$ limit and
saturates the anomalous Ward identities. We show the emergence of
a non-vanishing cosmological constant at $N\neq\infty$.
Simultaneously,  a
nonzero gluon condensate appears in the theory.
The vacuum degeneracy still holds. We analyze the
excitation spectrum of the model in Sect.~\ref{tre-finiteN-spectrum}.
The $\theta$-angle is then
included in the effective-Lagrangian  at finite $N$ (Sect.~\ref{tre-theta}).

Section \ref{quattro}   introduces a fermion mass term small with
respect to the confining scale of the theory. Here  we
show that a negative cosmological constant is generated by the
mass term,  and the
vacuum  degeneracy is lifted. A gluonic condensate linear in the fermion
mass turns on. Section~\ref{ntifm} treats the issue in the
limit $N \rightarrow \infty$.
The results thus obtained  are clearly
insensitive to the fermion representation.
Section~\ref{fnafm} is devoted to finite $m$ and finite $N$.
Since for $N=3$ the theory with
one antisymmetric fermion reduces to one-flavor QCD,  we
obtain, in this theory,  a
non-zero quark-antiquark condensate, a gluon  condensate and a
negative vacuum energy density, with and without the quark mass
term.

We find the scalar quark-antiquark meson to be heavier
than the  pseudoscalar one
(identified as ``$\eta^{\prime}$") in one-flavor QCD.

\section{Reviewing SYM  effective Lagrangian}
\label{due}

To warm up, before delving in the nonsupersymmetric case, it is
instructive to briefly review the construction of the SYM
effective Lagrangian. This will also give us an opportunity to
introduce relevant notation.

The fundamental Lagrangian of SU($N$) supersymmetric gluodynamics
is\,\footnote{The Grassmann integration is defined in such a way that
$\int \, \theta^2\, d^2\theta =2$.}
\be
{\cal L}& =&\frac{1}{4g^2}\,
\int \! {\rm d}^2 \theta\, \mbox{Tr}\,W^2 +{\rm H.c.} \nonumber\\[3mm]
& =&-\frac{1}{4 g^2} \,  G_{\mu\nu}^a G^{a\mu\nu}
+\frac{1}{2g^2}\, D^a D^a +\frac{i}{g^2} \lambda^a\sigma^\mu {\cal
D}_\mu\bar \lambda^a \, ,
\ee
where $g$ is the gauge coupling, the
vacuum angle is set to zero and \beq \mbox{Tr}\,W^2 \equiv
\frac{1}{2}W^{a,\alpha}W^{a}_{\alpha}= -
\frac{1}{2}\lambda^{a,\alpha}\lambda^{a}_{\alpha}\,. \eeq The
effective Lagrangian in supersymmetric gluodynamics was found by
Veneziano and Yankielowicz (VY) \cite{Veneziano:1982ah}. In terms
of the composite color-singlet chiral superfield $S$,
\beq S= \frac{3}{32\pi^2 N }\,\mbox{Tr}\,W^2 \,, \eeq
it can be written as follows:
 \be {\cal L}_{VY}&=& \frac{9\, N^2}{4\,\alpha}\, \int d^2\!\theta\,
d^2\!{\bar{\theta}}
\left(S S^{\dagger}\right)^{\frac{1}{3}} 
+\frac{N}{3} \int \! {\rm d}^2\theta\, \left\{ S  \ln
\left(\frac{S }{\Lambda^3}\right)^{N}-NS\right\}  + \mbox{H.c.} \,
, \label{VY} \nonumber \\\ee
where $\Lambda$ is a parameter related to the fundamental SYM
scale parameter. We singled out the factor $N^2$ in the K\"{a}hler
term to make the parameter $\alpha$ scale as $N^0$, see
Eq.~(\ref{alphascaling}) below. The standard definition of the
fundamental scale parameter is \cite{Hinchliffe}
\beq \Lambda_{\rm st} =\mu \left(\frac{16\pi^2}{\beta_0 \,
g^2(\mu)} \right)^{\beta_1/\beta_0^2} \exp\left( -
\frac{8\pi^2}{\beta_0 \, g^2 (\mu)} \right)\ ,
\eeq
which for SYM theory is exact \cite{nsvzbeta}
and reduces to
\beq \Lambda^3_{\rm SUSY\,\,YM} =\mu^3\left(\frac{16\pi^2
}{3N\,g^2(\mu)}\right) \exp\left( -\frac{8\pi^2}{N \, g^2 (\mu)}
\right)\,.
\eeq
The exact value of the gluino condensate was calculated by virtue
of holomorphy of SYM theory and is expressed in terms of
$\Lambda^3_{\rm SUSY\,\,YM}$ as follows
\cite{Shifman:1999mv,Davies:1999uw}:
\beq
 \langle S\rangle = \frac{9}{32\pi^2}\Lambda^3_{\rm SUSY\,\,YM}\,.
 \eeq
Comparing with Eq.~(\ref{VY}) we conclude that $\Lambda$ in
Eq.~(\ref{VY})
\beq \Lambda^3= \frac{9}{32\pi^2}\Lambda^3_{\rm SUSY\,\,YM}
\eeq
is $N$ independent. With our definitions, the gluino condensate
scales as $N$, in full accordance with the general rules.

Two remarks are in order here regarding the K\"ahler and
superpotential terms in the VY Lagrangian. First of all, the
K\"ahler term is ambiguous. The one presented in Eq.~(\ref{VY}) is
the simplest one compatible with symmetries of the theory. It
leads \cite{Kovner-Shifman}, however, to a spurious chirally
symmetric vacuum, which can be ruled out on general
grounds~\cite{Cachazo:2002ry}. This drawback is inessential for
our purposes.

Second, the logarithmic term of the superpotential in the
Lagrangian~(\ref{VY}) is not fully defined since the logarithm is
a multivalued function. The differences between the branches is
not important for the generation of the  anomalies --- this
difference resides in the invariant terms of ${\cal L}_{VY}$. The
proper definition was suggested in Ref.\ \cite{Kovner-Shifman}.
For the $n$-th branch one must define a corresponding Lagrangian,
\begin{equation}
{\cal L}_n=\frac{N}{3} \int \! {\rm d}^2\theta\,  S  \left[\ln
\left(\frac{S}{e\Lambda^3}\right)^{N} + 2i\pi n\right] +
\mbox{H.c.} \label{Ln} \,  ,
\end{equation}
where a specific branch is ascribed to  the logarithm. In terms of
the original theory the parameter $n$ shifts the vacuum angle
$\theta \to \theta+2\pi n$. The $Z_{N}$ invariance of the theory
is restored provided that the partition function sums over all
$n$,
\begin{equation} {\cal Z}=\sum_{n=-\infty}^\infty \int\!{\cal
D}\, S \,e^{\,i\!\int\! {\rm d}^4 x \,{\cal L}_n}\,.
\label{Lnprime}
\end{equation}
(The invariance group is $Z_{N}=Z_{2N}/Z_2$ because $S$ is
quadratic in $\lambda$, thus  identifying $\lambda$ and
$-\lambda$.) In what follows we will keep this in mind. What
remains to be added to fully specify the VY Lagrangian is the
normalization of the constant $\alpha$. This can be established by
requiring the mass of excitations to be $N$ independent,
\beq \alpha \sim {N^0} \,. \label{alphascaling}
\eeq
Indeed, the common mass of the bosonic and fermionic components of
$S$ is $M=2\alpha\, \Lambda /3$.  The chiral superfield $S$ at the
component level has the standard decomposition $S(y)=\varphi(y) +
\sqrt{2} \theta \chi(y) + \theta^2 F(y)$, where $y^\mu$ is the
chiral coordinate, $y^\mu=x^\mu - i \theta \sigma^\mu
\bar{\theta}$, and
\beq \varphi\ ,\quad \sqrt{2}\chi\ ,\quad F =\frac{3}{64\pi^2
N}\times \, \left\{
\begin{array}{l}
-\lambda^{a,\alpha}\lambda^{a}_{\alpha}\\[3mm]
G^a_{\alpha\beta}\lambda^{a,\beta} +2i D^a \lambda^{a}_{\alpha} \\[3mm]
-\frac{1}{2} G^a_{\mu\nu} G^{a\mu\nu}+
\frac{i}{2}G^a_{\mu\nu} \tilde{G}^{a\mu\nu}+\mbox{f.t.}
\end{array}
\right.
\label{decomp}
\eeq
where f.t. stands for (irrelevant) fermion terms.

The complex field $\varphi$ represents the scalar and pseudoscalar
gluino-balls while $\chi$ is their fermionic partner. It is
important that the $F$ field {\em must be treated as auxiliary}.
Although it is tempting to say that it represents the scalar and
pseudoscalar glueballs, treating $F$ as a dynamical field in the
minimal VY solution (\ref{VY}) is self-contradictory and leads to
paradoxes.

It should be stressed that the VY Lagrangian is not an effective
Lagrangian in the same sense as e.g. the pion chiral Lagrangian
which describes light degrees of freedom and can, therefore, be
systematically improved by introducing higher derivative terms.
The VY Lagrangian concisely summarizes the symmetries of the
underlying theory in terms of a ``minimal" number of degrees of
freedom whose choice is not unambiguous. Generalizations of the VY
Lagrangian containing more degrees of freedom were discussed in
the literature \cite{Farrar:1998rm,{Cerdeno:2003us}}. These
extensions were triggered, in part,  by lattice  simulations of
SYM spectrum \cite{Feo:2002yi}.

For our purposes of most importance are the scale and chiral
anomalies,
\begin{equation}
\partial^\mu J_\mu = \frac{N }{16\pi^2}\,
G_{\mu\nu}^a\tilde{G}^{a,\, \mu\nu}\, ,\qquad J_\mu= -
\frac{1}{g^2}\, \lambda^a \sigma_\mu \,\bar\lambda^a\, ,
\label{3anom}
\end{equation}
and
\beq \vartheta^\mu_{\mu}=- \frac{3 N }{32 \pi^2}\, G_{\mu\nu}^a
{G}^{a ,\, \mu\nu}\,\, ,
\eeq
where $J_\mu$ is the chiral current and $\vartheta^{\mu\nu}$ is
the standard (conserved and symmetric) energy-momentum tensor.

In SYM theory these two anomalies belong to the same
supermultiplet \cite{Ferrara-Zumino} and, hence, the coefficients
are the same (up to a trivial 3/2 factor due to normalizations).
In the orientifold theory the coefficients of the chiral and scale
anomalies coincide only at $N=\infty$; the subleading terms are
different.

Summarizing, the component form of the VY Lagrangian is
\beq {\cal
L}_{\rm VY}=\frac{N^2}{\alpha}\left(\varphi\, \bar\varphi
\right)^{-2/3}\,
\partial_\mu\bar\varphi\,\partial^\mu\varphi-\frac{4\, \alpha\,
N^2}{9}\, \left(\varphi\, \bar\varphi \right)^{2/3}
\ln\bar\varphi\,\ln\varphi +\mbox{fermions}\,, \label{vycomponent}
\eeq
where we set $\Lambda =1$ to ease the notation. The supersymmetry
of the bosonic part is expressed through (i) the K\"{a}hlerian
nature of the kinetic term; and (ii) the fact that the potential
term is the square of the absolute value of a holomorphic
function. In the limit $N\to\infty$ the same features should
persist in the orientifold theories. $1/N$ effects can destroy
either one or both.

\section{Effective Lagrangians in orientifold theories}
\label{tre}

Our task is to construct  the effective Lagrangian for the SU($N$)
gauge theory with one Dirac fermion in the two-index antisymmetric
representation of the gauge group. In this orientifold theory  the
trace and the chiral  anomalies are
\begin{eqnarray}
\vartheta^{\mu}_{\mu} &=&2N\left[N +
\frac{4}{9}\right]\left(F + \bar F\right) = -3\left[N +
\frac{4}{9}\right]\frac{1}{32\pi^2}\, G_{\mu\nu}^a {G}^{a ,\, \mu\nu} \ , \label{trace}\\[3mm]
\partial^{\mu} J_{\mu}
&=&i\,\frac{4N}{3}\,\left[ N - 2\right]\, \left(\bar F - F\right)=
\left[N - 2\right]\frac{1}{16\pi^2}\, G_{\mu\nu}^a {\tilde{G}}^{a
,\, \mu\nu} \ , \label{axial}
\end{eqnarray}
where
\begin{eqnarray}
\varphi= -\frac{3}{32\pi^2\, N}\,
 \widetilde{\psi}^{\alpha,[i,j]} \psi_{\alpha,[i,j]} \ ,
\label{phi1}
\end{eqnarray}
and $F$ is the same as in Eq.~(\ref{decomp}). The gluino field of
supersymmetric gluodynamics is replaced in the orientifold theory
by two Weyl fields, $ \widetilde{\psi}^{\alpha,[i,j]}$ and
$\psi_{\alpha,[i,j]}$, which can be combined into one Dirac
spinor. The color-singlet field $\varphi$ is now bilinear in $
\widetilde{\psi}^{\alpha,[i,j]}$ and  $\psi_{\alpha,[i,j]}$. Note
the absence of the color-singlet fermion field $\chi$ which was
present in supersymmetric gluodynamics. It no more exists in the
orientifold theory.

\subsection{Effective Lagrangian at $N\to\infty$ }

In this limit we can drop subleading $1/N$ terms in the
expressions for the trace and chiral anomaly. Then we will deal
with one and the same coefficient in both anomalies, much in the
same way as in SUSY gluodynamics. In fact, in this limit the boson
sector of the daughter theory is identical to that of the parent
one \cite{Armoni:2003gp}, and, hence,  the effective Lagrangian
must have  exactly the same form as in Eq. (\ref{vycomponent}),
with the omission of the fermion part and the obvious replacement
of $\lambda^a \lambda^a$ by $2\,
 \widetilde{\psi}^{\alpha,[i,j]} \psi_{\alpha,[i,j]}\,$
in the definition of $\varphi$.
It should be stressed again that the dynamical degrees described by
this Lagrangian are those related to $\varphi$, i.e. scalar and pseudoscalar
quark mesons. The field $F$ is non-dynamical;
therefore, the Lagrangian (\ref{vycomponent}) does {\em not}
implement Ward identities for gluonia.

Needless to say,  keeping the leading-$N$ terms only, we recover
all supersym\-metry-based bosonic properties such as degeneracy of
the opposite-parity mesons. Moreover, in this approximation the
vacuum energy vanishes.

\subsection{Effective Lagrangians in the orientifold
theories at finite $N$}
\label{tre-finite-N}

The most interesting question is that of subleading $1/N$ effects
which break the planar equivalence between the parent and daughter
theories. The effective Lagrangians approach might turn useful
since it is hard to compute $1/N$ corrections in the underlying
theory.

What changes must be introduced at finite $N$? First of all, the
overall normalization factor $N^2$ in Eq.~(\ref{vycomponent}) is
replaced by some function $f(N)$ such that $f(N)\to N^2$ at
$N\to\infty$. (One can also impose the condition
 $f(N)=0$ at $N=2$). Moreover, the anomalous
dimension of the operator $ \widetilde{\psi}^{\alpha,[i,j]}
\psi_{\alpha,[i,j]}\,$ no more vanishes. In fact, the
renormalization-group invariant combination is \beq
\left(g^2\right)^\delta\, \widetilde{\psi}^{\alpha,[i,j]}
\psi_{\alpha,[i,j]} \,,\qquad \delta \equiv
\frac{\left(1-\frac{2}{N}\right)
\left(1+\frac{1}{N}\right)}{\left(1+\frac{4}{9N}\right)}
-1\approx-\frac{13}{9N}\,. \eeq These effects affect certain
subtle details of the effective Lagrangian, but are unimportant in
its general structure. For the time being we will ignore them
focussing on the gross features of the $1/N$ corrections.

A constraint we impose is that we require (a) the finite-$N$
effective Lagrangian to pass into  the Lagrangian
(\ref{vycomponent}) once the $1/N$ corrections are dropped, and
(b) the scale and chiral anomalies (\ref{trace}), (\ref{axial}) to
be satisfied. The first requirement means, in particular, that we
continue to build ${\cal L}_{\rm eff}$ on a single (complex)
dynamical variable $\varphi$. Equations (\ref{trace}) and
(\ref{axial}) tell us that we cannot maintain the
``supersymmetric'' structure of the potentail term. We {\em have}
to ``untie'' the chiral and conformal dimensions of the fields in
the logarithms, see Eq. (\ref{vycomponent}). They cannot be just
powers of $\varphi$ since in this case the chiral and conformal
dimensions would be in one-to-one correspondence, and the
coefficients of the chiral and scale anomalies would be exactly
the same, modulo the normalization factor 3/2. At this stage a
non-holomorphicity must enter the game.

Let us introduce the fields
\beq \Phi =
\varphi^{1+\epsilon_1}\,\bar\varphi^{-\epsilon_2}\,,\qquad
\bar\Phi = \bar\varphi^{1+\epsilon_1}\,\varphi^{-\epsilon_2}\,,
\label{newfields} \eeq where $\epsilon_{1,2}$ are parameters
$O(1/N)$, \beq \epsilon_1 =- \frac{7}{9\,N}\,,\qquad \epsilon_2
=-\frac{11}{9\,N}\,. \label{newparam}
\eeq
The scale and chiral dimensions of $\bar\Phi$ and $\Phi$ are such
that using $\bar\Phi$ and $\Phi$ in the logarithms, we will solve
the problem of distinct $1/N$ corrections in the coefficients of
the scale  and chiral anomalies. The above replacement
(\ref{newfields}) is minimal in the sence that:

($\star$) it does not spoil the fact that $G^2$ and $G\tilde{G}$
are real and imaginary parts of a certain field.
(The operators $G^2$ and $G\tilde{G}$ will be identified
through the non-invariance of the Lagrangian under the scale
and chiral transformations, see below.)

This is not the end of the story, however. The vacuum expectation
value (VEV) of $G\tilde{G}$ vanishes in any gauge theory with
massless fermion fields. This is not the case with regards to the
vacuum expectation value of $G^2$, which, being proportional to
$\vartheta_\mu^\mu$, must develop a VEV at the subleading in $1/N$
level. Preserving the  property ($\star$)   above --- and we do
want to keep it --- leaves open  a single route: the $O(1/N)$ term
to be added to ${\cal L}_{\rm eff}$ which will give rise to
$\langle G^2\rangle$ must be scale invariant by itself.

As for the kinetic term, it was rather ambiguous even in the
supersymmetric case. It was constrained by the requirement of
scale and chiral invariance, and had to have a K\"{a}hlerian form.
$1/N$ corrections preserve the first requirement, while the second
one can be relaxed, generally speaking. To begin with, we will
make the simplest assumption and leave the kinetic term the same
as in ${\cal L}_{\rm VY}$. Other choices do not alter the overall picture
in the qualitative aspect.

Having said all that we can write down the effective Lagrangian
in the finite-$N$ orientifold theory,
\beq
 {\cal L}_{\rm eff}=f(N)\left\{
\frac{1}{\alpha}\left(\varphi\, \bar\varphi \right)^{-2/3}\,
\partial_\mu\bar\varphi\,\partial^\mu\varphi-\frac{4\alpha}{9}\,
\left(\varphi\, \bar\varphi \right)^{2/3}\,\left(
\ln\bar\Phi\,\ln\Phi - \beta \right)\right\}\,,
\label{fnocomponent}
\eeq
where $\beta $ is a numerical (real)
parameter,
\beq
\beta = O(1/N)\,, \eeq and \beq f(N) \to N^2
\,\,\, \mbox{at}\,\,\,  N\to\infty\,.
\eeq
The variations of this effective action under the scale and
chiral transformations
(i.e. $\varphi \to (1+3\gamma)\varphi $
and $\varphi \to (1+2i \gamma)\varphi $, respectively, with
real $\gamma$)
are
\begin{eqnarray}
\delta {\cal S}_{\rm eff}^{\rm scale}&=& \int d^4x\left\{-4\, \frac{\alpha\,
f}{3}\left(\varphi\, \bar\varphi \right)^{2/3}\left( 1+\epsilon_1
-\epsilon_2\right)
\left(\ln\bar\Phi +\ln \Phi\right)\right\}\,,\nonumber\\[4mm]
\delta {\cal S}_{\rm eff}^{\rm chiral}&=& \int d^4 x\left\{-8i\,\, \frac{\alpha\,
f}{9}\left(\varphi\, \bar\varphi \right)^{2/3}\left( 1+\epsilon_1
+\epsilon_2\right) \left(\ln\bar\Phi -\ln \Phi\right)\right\}\,,
\label{variat}
\end{eqnarray}
where the parameters $\epsilon_{1,2}$ are defined in Eq.~(\ref{newparam}).
Comparing with Eqs.~(\ref{trace}) and (\ref{axial})
we conclude that
\begin{eqnarray}
G_{\mu\nu}^a {G}^{a ,\, \mu\nu} &\propto& -N\left(\varphi\, \bar\varphi
\right)^{2/3}\left(\ln\bar\Phi +\ln \Phi\right)\,,\nonumber\\[3mm]
G_{\mu\nu}^a \tilde{G}^{a ,\, \mu\nu} &\propto& -N\,i\, \left(\varphi\, \bar\varphi
\right)^{2/3}\left(\ln\bar\Phi -\ln \Phi\right)\,.
\label{indenti}
\end{eqnarray}
Minimizing the potential term in the Lagrangian
(\ref{fnocomponent}) we find that the minimum occurs at \beq \ln
\varphi = \frac{2}{3}\,\beta +O(1/N^2)\,, \label{polmi} \eeq and
the minimal value of the potential energy --- i.e. the vacuum
energy density --- is
\beq V_{\rm min}={\cal E}_{\rm vac} = -\frac{4\alpha\,
f}{9}\,\beta + O(N^0)\,.
\label{vmin}
 \eeq
In this determination the value of ${\cal E}_{\rm vac}$ is
determined by the non-logarithmic term in the potential energy.
The logarithmic term enters only at the level $O(N^0)$.

There is a very important self-consistency check.
One can alternatively define the vacuum energy density as
$\frac{1}{4}\langle \vartheta^\mu_\mu\rangle $,
where the trace of the energy momentum tensor is in turn
proportional
to $G_{\mu\nu}^a {G}^{a ,\, \mu\nu}$, see Eq.~(\ref{indenti}).
In this method ${\cal E}_{\rm vac}$ will be determined exclusively
by the logarithmic term. In fact, it is not difficult to see that
\begin{eqnarray}
{\cal E}_{\rm vac} &=& -\frac{3N + {4}/{3}}{128\pi^2}\left\langle
G_{\mu\nu}^a
G^{a,\,\mu\nu}\right\rangle\nonumber\\[3mm]
&=& - \frac{\alpha\, f}{3}\,
\left\langle \ln\bar\Phi +\ln \Phi\right\rangle  + O(N^0) =-\frac{4\alpha\,
f}{9}\,\beta + O(N^0)\,,
\label{altt}
\end{eqnarray}
in complete agreement with Eq.~(\ref{vmin}).

\subsection{The sign of $\beta$}
\label{tsob}

{}From the above consideration it is clear that the (infrared part
of the) vacuum energy density is negative {\em if} $\beta >0$.
Here we would like to argue that this is indeed the case.

The first argument is phenomenological. At $N=3$ the theory we deal with
is in fact one-flavor QCD. In QCD with three light flavors
the gluon condensate is definitely positive while the
vacuum energy density is negative \cite{Shifman:bx}.
Actual QCD with three light flavors differs from
one-flavor QCD by making two quarks heavy.
Increasing the quark mass, somewhat changes the absolute value
of the gluon condensate but does not affect its sign
\cite{Novikov:xj}. Thus,
we expect  the (infrared part of the) vacuum energy density
to be negative in one-flavor QCD.

Another argument is based on the
implementation of the conformal Ward identities.
Let us forget about chiral Ward identities,
and try to build an (alternative) effective Lagrangian
which implements all (anomalous) scale Ward identities
in the correlation functions induced by the operator $G^2$,
with the field $\phi =G^2$ treated as dynamical rather than auxiliary.
This effective Lagrangian
has the form $ c\, G^2\,\ln G^2$, with  a positive constant $c$.
As was discussed in detail in Ref.~\cite{MS},
the vacuum energy density following from this
effective Lagrangian is necessarily negative.

\subsection{Lifting the spectrum degeneracy at finite $N$}
\label{tre-finiteN-spectrum}

At $N\to\infty$ the orientifold theory inherits from its
supersymmetric parent an infinite number of degeneracies in the
bosonic spectrum. At the effective Lagrangian level this property
manifests itself in the degeneracy of the scalar/pseudoscalar
mesons. At finite $N$ we expect this degeneracy to be lifted by
$1/N$ effects. To explore the  scalar/pseudoscalar splitting one
must study excitations near the vacuum in the Lagrangian
(\ref{fnocomponent}). Let us define
 \begin{eqnarray}
\varphi = \langle \varphi \rangle_{\rm vac}\left(1
 + a\,h\right)\ ,  \qquad
 h=\frac{1}{\sqrt{2}} \left(\sigma + i\,\eta^{\prime}
\right) \, ,
\end{eqnarray}
$\sigma$ and $\eta^{\prime}$ are two real fields and $a$ is a
constant which is determined by requiring the standard
normalization of the kinetic term for the complex field $h$,
\begin{eqnarray}
a^2=\frac{\alpha}{f} \, |\langle  \varphi \rangle|^{-\frac{2}{3}}
\ ,
\end{eqnarray}
Expanding around the ground state we find the masses of the scalar and the
pseudoscalar fields,
\begin{eqnarray}
 M_{\sigma} &=&\frac{2\alpha}{3}\, {\Lambda}
\left[1 + \epsilon_1 - \epsilon_2+ \frac{4}{9}\beta+ {
O}(N^{-2})\right]
\nonumber \\[3mm]
&=&\frac{2\alpha}{3}\, {\Lambda}\left[1 + \frac{4}{9N} +
\frac{4}{9}\beta+{O}(N^{-2})\right];
\nonumber \\[3mm]
M_{\eta^{\prime}} &=&\frac{2\alpha}{3}\, {\Lambda} \left[1 +
\epsilon_1 + \epsilon_2+{O}(N^{-2})\right]=\frac{2\alpha}{3}\,
{\Lambda} \left[1 - \frac{2}{N} +{O}(N^{-2})\right]  .
\label{spectrum}
 \end{eqnarray}
It is instructive  to consider the ratio of the pseudoscalar to scalar
mass,
 \begin{eqnarray}
\frac{M_{\eta^{\prime}}}{M_{\sigma}} = 1 +2\epsilon_2 -
\frac{4}{9}\beta + {O}(N^{-2}) =
 1 -\frac{22}{9N} -\frac{4}{9}\beta+ {
O}(N^{-2}) \ .\label{spectrum-ration}
\end{eqnarray}
It is clear that with positive $\beta$, corresponding to a
negative vacuum energy, the scalar state is heavier than the
its pseudoscalar counterpartner. As was mentioned, for $N=3$  the
two-index antisymmetric representation is equivalent to one-flavor QCD (with
a Dirac fermion in the fundamental representation). We
then predict that in one-flavor QCD the
scalar meson made   of one quark and one anti-quark is heavier
than the pseudoscalar one  (in QCD the latter is identified with the
$\eta^{\prime}$ meson).

\subsection{The $\theta$ angle in the orientifold field theories}
\label{tre-theta}

So far in our studies we have put the vacuum
angle $\theta =0$. Now we want to consider $\theta \neq 0$. The
vacuum angle is introduced, as usual, \beq {\cal L}_\theta
=\frac{\theta}{32\pi^2 }\, G_{\mu\nu}^a \tilde{G}^{a ,\,
\mu\nu}\,. \eeq Since the fermions in the orientifold theory under
consideration are massless (see Sect.~\ref{quattro} for a discussion
of a mass term) the vacuum angle is unobservable in  physical
quantities. It can be ``rotated away" by a rotation of the fermion
field. Correspondingly, the expression for the bifermion
condensate in the vacuum changes. In this subsection we will study
this change as well as the $\theta$ structure of the vacuum
family. Thus, to introduce the vacuum angle $\theta$ in the
effective Lagrangian (\ref{fnocomponent}) one must replace \beq
\varphi \to \varphi\, \exp\left(
-i\,\frac{\theta}{N-2}\right)\,,\qquad \Phi \to \Phi \exp\left(
-i\,\frac{\theta}{N}\right)\,. \eeq This is not the end of the
story, however, since the spontaneously broken $Z_{2(N-2)}$
symmetry (with the degeneracy of $N-2$ vacua which ensues) is not
properly implemented in Eq.~(\ref{fnocomponent}). The way to
implement it \cite{Kovner-Shifman} is outlined in Eqs.~(\ref{Ln}),
(\ref{Lnprime}). Namely, one must introduce an ``$n$-th
Lagrangian,"
\begin{eqnarray}
{\cal L}_{\rm eff}^{(n)}&=&f(N)\,N^{-2}\left\{ \frac{N^2}{\alpha}\left(\varphi\, \bar\varphi
\right)^{-2/3}\, \partial_\mu\bar\varphi\,\partial^\mu\varphi\right.
\nonumber\\[3mm]
&-&\left. \frac{4\alpha}{9}\, \left(\varphi\, \bar\varphi
\right)^{2/3}\,\left[ \left|\ln\bar\Phi^N +2\pi i
n+i\theta\right|^2\,
 - N^2\beta \right]\right\}\,,
\end{eqnarray}
where $n$ labels the
branches  of  the logarithm. In terms of the original
theory the parameter $n$ shifts the vacuum angle $\theta \to \theta+2\pi n$.
The proper
$Z_{2(N-2)}$ invariance of the theory
 is restored provided that the partition function sums over
all $n$,
\begin{equation}
{\cal Z}=\sum_{n=-\infty}^\infty \int\!{\cal D}\, S \,e^{\,i\!\int\! {\rm d}^4 x
\,{\cal L}_n}\,.
\label{pfsums}
\end{equation}

\subsection{Orientifold theory with fermions in the
two-index symmetric representation}
\label{tisym}

 Here the trace
and axial anomalies are
\begin{eqnarray}
\vartheta^{\mu}_{\mu} &=&2N\left[N  - \frac{4}{9}\right]\left(F +
\bar F\right) = -3\left[N -
\frac{4}{9}\right]\frac{1}{32\pi^2}\, G_{\mu\nu}^a {G}^{a ,\, \mu\nu} \ , \label{traces}\\[3mm]
\partial^{\mu} J_{\mu}
&=&i\,\frac{4N}{3}\,\left[ N + 2\right]\, \left(\bar F - F\right)=
\left[N + 2\right]\frac{1}{16\pi^2}\, G_{\mu\nu}^a {\tilde{G}}^{a
,\, \mu\nu} \ , \label{axials}
\end{eqnarray}
where $F$ is given in Eq.~(\ref{decomp}). The gluino field of
supersymmetric gluodynamics is replaced in this orientifold theory
by two Weyl fields,  $ \widetilde{\psi}^{\alpha,\{i,j\}}$ and
$\psi_{\alpha,\{i,j\}}$ and the color-singlet field $\varphi$ is
now
\begin{eqnarray}
\varphi= -\frac{3}{32\pi^2\, N}\,
 \widetilde{\psi}^{\alpha,\{i,j\}} \psi_{\alpha,\{i,j\}} \ .
\label{phi2}
\end{eqnarray}
In the $N\rightarrow \infty$ limit the low-energy effective
Lagrangian again corresponds to the bosonic part of the VY
Lagrangian. To construct the effective Lagrangian at finite $N$ we
follow the same procedure used when the two fermions were in the
two-index antisymmetric representation. To take into account
non-holomorphicity we introduce the fields
\beq \Phi =
\varphi^{1+\hat{\epsilon}_1}\,\bar\varphi^{-\hat{\epsilon}_2}\,,\qquad
\bar\Phi =
\bar\varphi^{1+\hat{\epsilon}_1}\,\varphi^{-\hat{\epsilon}_2}\,,
\label{newfieldsS} \ .
\eeq
The coefficients of the trace and axial anomaly again fix
$\hat{\epsilon}_{1,2}$,
\beq \hat{\epsilon}_1 = \frac{7}{9\,N}\,,\qquad \hat{\epsilon}_2
=\frac{11}{9\,N}\ . \label{newparam2} \eeq
The effective Lagrangian in the finite-$N$ orientifold theory is:
\beq {\cal L}_{\rm eff}=f(N)\left\{
\frac{1}{\alpha}\left(\varphi\, \bar\varphi \right)^{-2/3}\,
\partial_\mu\bar\varphi\,\partial^\mu\varphi-\frac{4\alpha}{9}\,
\left(\varphi\, \bar\varphi \right)^{2/3}\,\left(
\ln\bar\Phi\,\ln\Phi - \hat{\beta} \right)\right\}\,,
\label{fnocomponentS} \eeq
where $\hat{\beta} $ is a numerical (real) parameter $\hat{\beta}
= O(1/N)$ and $f(N) \to N^2$ at $N\to\infty$. The potential term
in the Lagrangian (\ref{fnocomponentS}) is minimized for
\beq \ln \varphi = \frac{2}{3}\,\hat{\beta} +O(1/N^2)\,,
\label{polmiS}
\eeq
and the minimal value of the potential energy
--- i.e. the vacuum energy density
--- is
\beq V_{\rm min}= -\frac{4\alpha\, f}{9}\,\hat{\beta} + O(N^0)\,.
\label{vminS} \eeq
At finite $N$ the spectrum degeneracy is lifted by $1/N$
corrections. Expanding around the ground state
we obtain that the masses of the
scalar and the pseudoscalar fields assume the same form as for the
two-index antisymmetric representation (see (\ref{spectrum})) with
$\epsilon$ and $\beta$ replaced by $\hat{\epsilon}$ and
$\hat{\beta}$. The mass ratio between the pseudoscalar and the
scalar particles in this theory is
 \begin{eqnarray}
\frac{M_{\eta^{\prime}}}{M_{\sigma}} = 1 +2\hat{\epsilon}_2 -
\frac{4}{9}\hat{\beta} + { O}(N^{-2}) =
 1 +\frac{22}{9N} -\frac{4}{9}\hat{\beta}+ {
O}(N^{-2}) \ .
\end{eqnarray}
Interestingly,  if $\hat{\beta}$ is sufficiently small (and is
positive, see above), the scalar meson could be lighter
 than the corresponding pseudoscalar one.

The $\theta$ dependence in the effective Lagrangian
(\ref{fnocomponentS}) is obtained by replacing \beq
 \varphi \to
\varphi\, \exp\left( -i\,\frac{\theta}{N+2}\right)\,,\qquad \Phi
\to \Phi \exp\left( -i\,\frac{\theta}{N}\right)\,. \eeq
To properly implement the spontaneously broken $Z_{2(N+2)}$
symmetry at the effective Lagrangian level \cite{Kovner-Shifman}
we write
\begin{eqnarray}
{\cal L}_{\rm eff}^{(n)}&=&f(N)\,N^{-2}\left\{
\frac{N^2}{\alpha}\left(\varphi\, \bar\varphi \right)^{-2/3}\,
\partial_\mu\bar\varphi\,\partial^\mu\varphi\right.
\nonumber\\[3mm]
&-&\left. \frac{4\alpha}{9}\, \left(\varphi\, \bar\varphi
\right)^{2/3}\,\left[ \left|\ln\bar\Phi^N +2\pi i
n+i\theta\right|^2\,
 - N^2\hat{\beta} \right]\right\}\,,
\end{eqnarray}
where $n$ labels the branches  of  the logarithm,
and require  the partition function to sum over
all $n$, as in Eq.~(\ref{pfsums}).

\section{Adding a mass term}
\label{quattro}

The most straightforward approach is to add a ``soft''
supersymmetry breaking term to the Lagrangian. This was carried
out by Masiero and Veneziano \cite{Masiero-Veneziano} who
introduced a gluino mass term,
\begin{eqnarray}
\Delta {\cal L}_{m} = -\frac{m}{2g^2} \, \lambda \lambda + {\rm
h.c.} \ .
\end{eqnarray}
which at the effective-Lagrangian level translates as
\begin{equation}
\Delta {\cal L}_{m}=\frac{m}{g^2}
\frac{N}{3}\,32 \pi^2\,\left(\varphi + \bar{\varphi}\right) =
\frac{4\, m}{3\lambda}\, N^2 \left(\varphi + \bar{\varphi}\right)\,,
\label{soft}
\end{equation}
where we introduced the 't Hooft coupling
\beq
\lambda\equiv \frac{g^2 N}{8\pi^2}\,.
\eeq
It is convenient to assume the mass parameter $m$ to be
real and positive. One can always make it
real and positive at the price of redefining the vacuum angle
$\theta$. In what follows we will adopt this convention.

The softness restriction is $m /\lambda  \ll {\Lambda}$.
Recently, soft SUSY breaking has been reanalyzed in \cite{EHS}, while
a model for not-soft breaking has been proposed in
\cite{Sannino:1997dd}.

Note that the combination $m/\lambda$ is
renormalization-group invariant to leading order, and scales as
$N^0$; the one which is renormalization-group invariant to all
orders can be found too, see \cite{Hisano:1997ua}.  Analysis
 of this model indicates that the theory is
``trying'' to approach the non-SUSY Yang-Mills case. Namely, the spin-$0$
and spin-$1/2$ particles split from each other, and their masses
each pick up a piece linear in $m$. One of   $N$ distinct vacua
of the SUSY theory
becomes the true minimum. Furthermore, the vacuum value of the
gluon  condensate is no longer zero.

Analogously, we will study now    non-zero mass effects for
the fermions in the orientifold theory. The mass term    at the level of the
underlying theory  is represented by
\begin{eqnarray}
\Delta {\cal L}_{m}= -\frac{m}{g^2}\,\psi \widetilde{\psi} + {\rm h.c.} \ .
\end{eqnarray}
The sum over  spin is understood, and the color indices are
contracted in order to construct a color singlet for the theories
under consideration.

\subsection{$N\to\infty$, finite $m$}
\label{ntifm}

In the large-$N$ limit  the mass term is blind with regards
to the representation to which the underlying fermions belong.
At the effective-Lagrangian level, for small $m$, we add the
term (\ref{soft}) to the Lagrangian (\ref{vycomponent}).
Expanding in $m$ around the $N=\infty$ solution we find, at
 leading order
\begin{eqnarray}
\langle  \varphi \rangle \simeq {\Lambda}^3 \left(1 + \delta
\right) \ , \qquad {\rm with} \qquad \delta = \frac{3 \, m
 }{\alpha\, \lambda\,\Lambda} \ .
\end{eqnarray}
This induces a VEV for  the gluon condensate. In the presence of a
non zero fermion mass term the trace of the energy momentum tensor
reads:
\begin{eqnarray}
\vartheta^{\mu}_{\mu}&=&-\frac{3}{32\pi^2}\left[N+\frac{4}{9}\right]
G^a_{\mu\nu}G^{a,\mu\nu}+\frac{m}{g^2}\left(\psi
\widetilde{\psi} + {\rm h.c.}\right) \nonumber
\\&\simeq&-\frac{3N}{32\pi^2}G^a_{\mu\nu}G^{a,\mu\nu}-
\frac{4m}{3\lambda}N^2\left(\varphi + \bar{\varphi}\right) \ .
\end{eqnarray}
Since $\langle \vartheta^{\mu}_{\mu} \rangle/4= {\cal E}_{\rm
vac}$:
\begin{eqnarray}
 \frac{\langle G^a_{\mu\nu}G^{a,\mu\nu} \rangle}{64\pi^2} =
\frac{4\, N \, m}{3\lambda}\, \langle \varphi  \rangle + {
O}\left(m^2\right)\ ,
\end{eqnarray}
exhibiting the beginning of the formation of the gluon condensate
as a fermion mass term is introduced. The same result can be found
using the equation of motion for the auxiliary field $F$
\cite{Masiero-Veneziano}.

The degeneracy between the scalar and pseudoscalar mesons is
lifted too \cite{Masiero-Veneziano,{EHS}}. The vacuum energy
density takes the form
\begin{eqnarray}
{\cal E}_{\rm vac}= V_{\rm min} =-
\frac{8N^2}{3\,\lambda}\, m \Lambda^3 +O(m^2)\,,
\label{massvacuum}
\end{eqnarray}
which is explicitly proportional to $N^2$ and $m$. If the
$\theta$-angle dependence is added the vacuum energy modifies as
follows:
\begin{eqnarray} {\cal E}_{\rm vac} =
\frac{8N^2}{3\,\lambda}\, m \Lambda^3\, {\rm
min}_{k}\left\{-\cos\left[\frac{\theta + 2\pi\,k
}{N}\right]\right\} .
\end{eqnarray}
The $N$-fold degeneracy of the vacuum is lifted, and  one ends up
with a unique vacuum state \cite{Masiero-Veneziano}.

\subsection{Finite $N$ and $m$}
\label{fnafm}

The leading $1/N$ and $m$ corrections can be considered
simultaneously using the Lagrangian:
\beq f(N)\left\{ \frac{1}{\alpha}\left(\varphi\, \bar\varphi
\right)^{-2/3}\,
\partial_\mu\bar\varphi\,\partial^\mu\varphi-\frac{4\alpha}{9}\,
\left(\varphi\, \bar\varphi \right)^{2/3}\,\left(
\ln\bar\Phi\,\ln\Phi - \beta \right)\right\} + \frac{4\,
m}{3\lambda}\, N^2 \left(\varphi + \bar{\varphi}\right)\ .
\label{fnocomponent+mass} \eeq
The vacuum expectation value reads:
\begin{eqnarray}
\langle \varphi \rangle = \Lambda^3 \left(1 + \frac{2}{3}\beta +
\frac{3m}{\alpha \lambda \Lambda}\right ) + {
O}\left(m^2,N^{-2},mN^{-1}\right)\ ,
\end{eqnarray}
yielding the following vacuum energy density:
\begin{eqnarray}
{\cal E}_{\rm vac}=V_{\rm min} = -\frac{4\alpha f}{9} \beta
\Lambda^4 - \frac{8N^2}{3\lambda} m\Lambda^3 +{
O}\left(m^2,N^{0},mN\right) \ .
\end{eqnarray}
For the spectrum we have:
\begin{eqnarray}
 M_{\sigma} &=&
\frac{2\alpha}{3}\, {\Lambda}\left[1 + \frac{4}{9N} +
\frac{4}{9}\beta+ \frac{5}{2}\frac{m}{\alpha \lambda \Lambda}+
{O}(m^2,N^{-2},mN^{-1})\right];
\nonumber \\[3mm]
M_{\eta^{\prime}} &=&
\frac{2\alpha}{3}\, {\Lambda} \left[1 - \frac{2}{N} +
\frac{3}{2}\frac{m}{\alpha \lambda \Lambda}+
{O}(m^2,N^{-2},mN^{-1})\right] , \label{spectrumNm}
 \end{eqnarray}
and the following is the new ratio of the pseudoscalar to scalar
mass,
 \begin{eqnarray}
\frac{M_{\eta^{\prime}}}{M_{\sigma}} = 
 1 -\frac{22}{9N} -\frac{4}{9}\beta -  \frac{m}{\alpha \lambda \Lambda}+
{O}(m^2,N^{-2},mN^{-1}) \ .\label{spectrum-ration-Nm}
\end{eqnarray}
The gluon condensate becomes:
\begin{eqnarray}
 \frac{\langle G^a_{\mu\nu}G^{a,\mu\nu} \rangle}{64\pi^2} =
\frac{4\, N \, m}{3\lambda}\, \Lambda^3 +
\frac{8}{27}\alpha\,N\beta\Lambda^4+ { O}\left(m^2,N^{-1},mN^0
\right)\ .
\end{eqnarray}
These results show that the contribution of the fermion mass
reinforces the effect of the finite $N$ contribution.
Interestingly the scalar state becomes even more massive than the
pseudoscalar state when considering finite both $N$ and $m$.

The $\theta$-angle dependence of the vacuum energy for the
fermions in the two-index antisymmetric representation of the
gauge group is
\begin{eqnarray} {\cal E}_{\rm vac} =
\frac{8N^2}{3\,\lambda}\, m \Lambda^3\, {\rm
min}_{k}\left\{-\cos\left[\frac{\theta + 2\pi\,k
}{N-2}\right]\right\}  -\frac{4\alpha f}{9} \beta \Lambda^4 \
.\end{eqnarray}
The $N-2$-fold vacuum degeneracy is lifted due to the presence of
a mass term in the theory, yielding a unique vacuum.

\section{Conclusions}
\label{sec:Discussion}

We constructed  the effective Lagrangians of the
Veneziano-Yankielowicz type for orientifold field
theories, starting from the underlying  SU$(N)$
gauge theory with the Dirac fermion in the two-index antisymmetric
(symmetric)
representation of the gauge group.
These Lagrangians
incorporate ``important" low-energy degrees of freedom (color singlets)
and implement (anomalous) Ward identities. At $N\to\infty$
they coincide with the bosonic part of the VY Lagrangian.
 The orientifold effective
Lagrangians at $N=\infty$ display the vanishing of the cosmological constant
and the spectral  degeneracy (i.e. the scalar-pseudoscalar
degeneracy).

The most interesting question we addressed is the
finite-$N$/finite-$m$ generalization.
To the leading order in $1/N$ we demonstrated the occurrence of a
negative vacuum energy density, and of the gluon  condensate.
We first derived these results at $m=0$ and then
extended them to include the case $m\neq 0$.

At $N=3$ the theory with one  Dirac
fermion in the two-index antisymmetric
representation of the gauge group is in fact one-flavor QCD.
Our analysis of the finite-$N$ effective Lagrangian
illustrates the emergence of the gluon
condensate in this theory.
The vacuum degeneracy typical of
supersymmetric gluodynamics does not disappear
at finite $N$. However, introduction of
mass $m\neq 0$ lifts the vacuum degeneracy,
in full compliance with the previous expectations.
Both effects, $N\neq \infty$ and $m\neq 0$ conspire
to get lifted
 the scalar-pseudoscalar
degeneracy. We evaluated the ratio $M_{\eta^{\prime}}/M_{\sigma}$.

Finally, we studied the effects of finite $\theta$, as they are
exhibited in the orientifold effective Lagrangian. Here our
conclusions can be summarized as follows.

(i) We have shown that at finite $N$ and $m=0$
 a negative vacuum energy density,  together
with the gluon condensate, appear  in the orientifold theories under
consideration. The $(N-2)$ vacuum degeneracy is unaffected by
$1/N$ corrections.
Physical quantities (such as  the vacuum energy) do not depend on
$\theta$;

(ii) A non-zero mass term   lifts the vacuum degeneracy.
Simultaneously,  the vacuum energy density acquires a $\theta$
dependence. In terms of the effective Lagrangian, this $\theta$
dependence requires a Lagrangian glued from $N-2$ pieces, as in
Eq.~(\ref{pfsums}).

In conclusion, we reiterate that the effective Lagrangian
obtained in the present work summarizes dynamics of
a single degree of freedom, the complex field $\varphi \propto \lambda\lambda$
(and its complex conjugate). The degrees of freedom associated with
$G^2$ and $G\tilde G$ are non-dynamical.
As the value of the gluino mass $m$ grows,  degrees of freedom associated with
$\lambda\lambda$ become heavier. At $m\sim\Lambda$ they loose their status
of ``important low-energy degrees of freedom." In fact, at larger $m$ they can be integrated out. Therefore, our construction is valid only at
$m\ll\Lambda$. In the opposite limit  the traces of supersymmetry
disappear, the ``important low-energy degrees of freedom"
are those associated with $G^2$ and $G\tilde G$,
and one gets a totally different effective Lagrangian
built on the fields $G^2$ and $G\tilde G$ as dynamical variables.

\Acknowledgements

\noindent We would like to thank A. Armoni and
G. Veneziano for thorough and insightful discussions. The initial
stages of this work were carried out at CERN. We are  grateful to
CERN Theory Division for kind hospitality.

The work of F.S. is
supported by the Marie--Curie Foundation under contract
MCFI-2001-00181. The work of M.S. is supported in part  by DOE
grant DE-FG02-94ER408.


\begin{thebibliography}{9}

\bibitem{Armoni:2003gp}
A.~Armoni, M.~Shifman and G.~Veneziano,
Nucl.\ Phys.\ B {\bf 667}, 170 (2003)
[hep-th/0302163].

\bibitem{Armoni:2003fb}
A.~Armoni, M.~Shifman and G.~Veneziano,
{\em SUSY relics in one-flavor QCD from a new $1/N$ expansion,}
Phys. Rev. Lett., to appear [hep-th/0307097].

\bibitem{Shifman:1999mv}
M.~A.~Shifman and A.~I.~Vainshtein,
{\em Instantons versus supersymmetry: Fifteen years later,}
in M. Shifman, {\em ITEP Lectures on Particle Physics and Field Theory},
(World Scientific, Singapore, 1999)
Vol. 2, p. 485-647
[hep-th/9902018].

\bibitem{Shifman:ia}
M.~A.~Shifman and A.~I.~Vainshtein,
Nucl.\ Phys.\ B {\bf 296}, 445 (1988).

\bibitem{ASV3}
A.~Armoni, G. Veneziano and M.~Shifman,  {\em QCD Quark Condensate
from SUSY and the Orientifold Large-$N$ Expansion,}
hep-th/0303109.

\bibitem{Armoni:1999gc}
A.~Armoni and B.~Kol,
JHEP {\bf 9907}, 011 (1999) [hep-th/9906081].

\bibitem{Angelantonj:1999qg}
C.~Angelantonj and A.~Armoni,
Nucl.\ Phys.\ B {\bf 578}, 239 (2000) [hep-th/9912257].

\bibitem{schechter}
J.~Schechter,
Phys.\ Rev.\ D {\bf 21} (1980) 3393.

\bibitem{joe}
C.~Rosenzweig, J.~Schechter and G.~Trahern, Phys. Rev. {\bf D21},
3388 (1980); P.~Di Vecchia and G.~Veneziano, Nucl. Phys. {\bf
B171}, 253 (1980); E.~Witten, Ann. of Phys. {\bf 128}, 363 (1980);
P.~Nath and A.~Arnowitt, Phys. Rev. {\bf D23}, 473 (1981);
A.~Aurilia, Y.~Takahashi and D.~Townsend, Phys. Lett. {\bf 95B},
65 (1980); K.~Kawarabayashi and N.~Ohta, Nucl. Phys. {\bf B175},
477 (1980).

\bibitem{MS}
A.~A.~Migdal and M.~A.~Shifman,
Phys.\ Lett.\ B {\bf 114}, 445 (1982);
J.~M.~Cornwall and A.~Soni,
Phys.\ Rev.\ D {\bf 29}, 1424 (1984);
Phys.\ Rev.\ D {\bf 32}, 764 (1985).

\bibitem{SST}
A.~Salomone, J.~Schechter and T.~Tudron, Phys. Rev. {\bf D23},
1143 (1981); J.~Ellis and J. Lanik, Phys. Lett. {\bf 150B}, 289
(1985); H.~Gomm and J.~Schechter, Phys. Lett. {\bf 158B}, 449
(1985); F.~Sannino and J.~Schechter,
Phys.\ Rev.\ D {\bf 60}, 056004 (1999) [hep-ph/9903359].

\bibitem{Veneziano:1982ah}
G.~Veneziano and S.~Yankielowicz,
Phys.\ Lett.\ B {\bf 113}, 231 (1982).

\bibitem{Sannino:2002wb}
F.~Sannino,
Phys.\ Rev.\ D {\bf 66}, 034013 (2002) [hep-ph/0204174].

\bibitem{Mocsy:2003tr}
A.~M\'ocsy, F.~Sannino and K.~Tuominen,  Phys.\ Rev.\ Lett.\  {\bf 91}, 092004
(2003) [hep-ph/0301229].

\bibitem{Mocsy:2003qw}
A.~M\'ocsy, F.~Sannino and K.~Tuominen,
{\em Confinement versus chiral symmetry,}
hep-ph/0308135.

%

\bibitem{Hinchliffe}
I. Hinchliffe, Phys. Rev. {\bf D66}, \# 1-I, 010001-89 (2002).

\bibitem{nsvzbeta}
V.~A.~Novikov, M.~A.~Shifman, A.~I.~Vainshtein and V.~I.~Zakharov,
Nucl.\ Phys.\ B {\bf 229}, 381 (1983);
Phys.\ Lett.\ B {\bf 166}, 329 (1986).

\bibitem{Davies:1999uw}
N.~M.~Davies, T.~J.~Hollowood, V.~V.~Khoze and M.~P.~Mattis,
 Nucl.\ Phys.\ B {\bf 559}, 123 (1999)
[hep-th/9905015]. Note  that an unconventional definition of the
scale parameter $\Lambda$ is used in Ref.~\cite{Davies:1999uw}.
One can pass to the conventional definition of $\Lambda$ either by
normalizing the result to the SU(2) case \cite{Shifman:1999mv} or
by analyzing the context of Ref.~\cite{Davies:1999uw}. Both
methods give the same result.

\bibitem{Kovner-Shifman}
A.~Kovner and M.~A.~Shifman,  Phys.\ Rev.\ D {\bf 56}, 2396 (1997)
[hep-th/9702174].

\bibitem{Cachazo:2002ry}
F.~Cachazo, M.~R.~Douglas, N.~Seiberg and E.~Witten,  JHEP {\bf
0212}, 071 (2002) [hep-th/0211170].

\bibitem{Farrar:1998rm}
G.~R.~Farrar, G.~Gabadadze and M.~Schwetz,
Phys.\ Rev.\ D {\bf 60}, 035002 (1999) [hep-th/9806204].

\bibitem{Cerdeno:2003us}
D.~G.~Cerdeno, A.~Knauf and J.~Louis, {\em A note on effective
${\cal N} = 1$ super Yang-Mills theories versus lattice results,}
hep-th/0307198.

\bibitem{Feo:2002yi}
I.~Montvay,
Nucl.\ Phys.\ Proc.\ Suppl.\  {\bf 63}, 108 (1998)
[hep-lat/9709080];
N.~Evans, S.~D.~Hsu and M.~Schwetz,
hep-th/9707260.
A.~Donini, M.~Guagnelli, P.~Hernandez and A.~Vladikas,
Nucl.\ Phys.\ B {\bf 523}, 529 (1998) [hep-lat/9710065];
For a recent review see A.~Feo, {\em Supersymmetry on the
lattice,} hep-lat/0210015.

\bibitem{Ferrara-Zumino}
S.~Ferrara and B.~Zumino, Nucl. Phys. {\bf B87}, 207 (1975).

\bibitem{Shifman:bx}
M.~A.~Shifman, A.~I.~Vainshtein and V.~I.~Zakharov,
Nucl.\ Phys.\ B {\bf 147}, 385; 448 (1979).

\bibitem{Novikov:xj}
V.~A.~Novikov, M.~A.~Shifman, A.~I.~Vainshtein and V.~I.~Zakharov,
Nucl.\ Phys.\ B {\bf 191}, 301 (1981).

\bibitem{Masiero-Veneziano}
A.~Masiero and G.~Veneziano, Nucl. Phys. {\bf B249}, 593 (1985).

\bibitem{EHS}
N.~Evans, S.D.H.~Hsu and M.~Schwetz, Phys. Lett. B {\bf 404}, 77
(1997); Nucl. Phys. {\bf B484}, 124 (1997); N.~Evans, S.D.H.~Hsu,
M.~Schwetz, S.B.~Selipsky, Nucl. Phys. {\bf B456}, 205 (1995).

\bibitem{Sannino:1997dd}
F.~Sannino and J.~Schechter,
Phys.\ Rev.\ D {\bf 57}, 170 (1998) [hep-th/9708113];
S.~D.~Hsu, F.~Sannino and J.~Schechter,
Phys.\ Lett.\ B {\bf 427}, 300 (1998) [hep-th/9801097].

\bibitem{Hisano:1997ua}
J.~Hisano and M.~A.~Shifman,
Phys.\ Rev.\ D {\bf 56}, 5475 (1997) [hep-ph/9705417].

\end{thebibliography}
\end{document}